\documentclass[aps,prmat,nobibnotes,nofootinbib,superscriptaddress,twocolumn,
longbibliography
]{revtex4-1}

\usepackage{graphicx} 	
\usepackage{dcolumn} 	
\usepackage{bm}   		
\usepackage{amssymb}   	
\usepackage[separate-uncertainty=true,range-units=single,range-phrase=--,retain-explicit-plus=true]{siunitx} 
\usepackage[svgnames]{xcolor}	
\usepackage[version=3]{mhchem}	

\usepackage{xr} 
\usepackage[colorlinks,
citecolor=Green,urlcolor=RoyalBlue]{hyperref}
\usepackage[caption=false]{subfig}
\usepackage[nameinlink,capitalise]{cleveref}
\usepackage{lipsum}

\newcommand{\ch}{$C_{\ce{H2}}$}
\DeclareSIUnit\unitcell{uc}

\graphicspath{{figs/}}

\newif\ifsuppinpaper \suppinpapertrue

\ifsuppinpaper
\else
\fi

\def\Tim{T_\mathrm{IM}}
\def\Tmi{T_\mathrm{MI}}
\def\Tp{T_\mathrm{P}}
\def\Th{T_\mathrm{h}}
\def\Tsm{T_\mathrm{sm}}
\def\Tr{T_\mathrm{raw}}
\def\Tins{T_\mathrm{IR,\,ins}}
\def\Tmet{T_\mathrm{IR,\,met}}
\def\Ai{A_\mathrm{ins}}
\def\jp{j_\mathrm{peak}}
\def\ja{j_\mathrm{all\text{-}met}}

\begin{document}
\title{Role of local temperature in the current-driven metal--insulator transition of \ce{Ca2RuO4}}

\author{Giordano Mattoni}
\email{mattoni@scphys.kyoto-u.ac.jp}
\affiliation{Department of Physics, Graduate School of Science, Kyoto University, Kyoto 606-8502, Japan}

\author{Shingo Yonezawa}
\affiliation{Department of Physics, Graduate School of Science, Kyoto University, Kyoto 606-8502, Japan}

\author{Fumihiko Nakamura}
\affiliation{Department of Education and Creation Engineering, Kurume Institute of Technology, Fukuoka 830-0052, Japan}

\author{Yoshiteru Maeno}
\affiliation{Department of Physics, Graduate School of Science, Kyoto University, Kyoto 606-8502, Japan}

\begin{abstract}

It was recently reported that a continuous electric current is a powerful control parameter to trigger changes in the electronic structure and metal--insulator transitions (MITs) in \ce{Ca2RuO4}.
However, the spatial evolution of the MIT and the implications of the unavoidable Joule heating have not been clarified yet, often hindered by the difficulty to asses the local sample temperature.
In this work, we perform infrared thermal imaging on single-crystal \ce{Ca2RuO4} while controlling the MIT by electric current.
The change in emissivity at the phase transition allows us to monitor the gradual formation and expansion of metallic phase upon increasing current.
Our local temperature measurements indicate that, within our experimental resolution, the MIT always occurs at the same local transition temperatures, irrespectively if driven by temperature or by current.
Our results highlight the importance of local heating, phase coexistence, and microscale inhomogeneity when studying strongly correlated materials under the flow of electric current.

\end{abstract}

\date{\today}
\maketitle

\section{Introduction}

In the field of quantum materials, extensive research interest has been recently focusing on strongly correlated systems, where unique changes of physical properties can be induced by external stimuli \cite{
	keimer2017physics,
	basov2017towards
}.
A paradigmatic example is the layered perovskite \ce{Ca2RuO4}, a material having a Mott insulating ground state with antiferromagnetic ordering \cite{
	nakatsuji1997ca2ruo4,
	braden1998crystal,
	alexander1999destruction
}.
In this material, exotic phases presenting orbital ordering and unconventional magnetic modes have been reported \cite{
	zegkinoglou2005orbital,
	jain2017higgs
}.
Metallic states with paramagnetic, ferromagnetic, or superconducting character have been triggered by a series of control parameters, such as chemical composition \cite{
	cao2000ground,
	carlo2012new
}, pressure \cite{
	nakamura2002mott,
	steffens2005high,
	alireza2010evidence
}, or strain \cite{
	miao2013epitaxial,
	shukla2016electrically,
	dietl2018tailoring,
	ricco2018situ
}.

More recently, it has been reported that a continuous electric current induces structural and electronic changes in \ce{Ca2RuO4} \cite{
	nakamura2013electric,
	okazaki2013current
}.
Several subsequent works reported the emergence of novel phases in these non-equilibrium steady states induced by current \cite{
	bertinshaw2019unique,
	cirillo2019emergence,
	zhao2019nonequilibrium,
	okazaki2020current
}.
An interesting development is the possibility of using a continuous current to trigger the material metal--insulator transition (MIT) \cite{
	nakamura2013electric
}, which naturally occurs at about \SI{360}{\kelvin} in bulk single crystals \cite{
	alexander1999destruction
}.
In this framework, coexistence of metallic and insulating regions stabilised by current have been reported by Zhang \textit{et al.}, with a phase front characterised by nanoscale stripes of alternating phases \cite{
	zhang2019nano
}.

However, the considerable amount of power supplied by a flowing current requires one to carefully take into account the unavoidable Joule heating, which makes it often difficult to measure the actual sample temperature.
Recent reports by F\"{u}rsich \textit{et al.} indicate that heating by direct current can lead to a large temperature rise in \ce{Ca2RuO4}, strongly depending on the cooling technique \cite{
	fursich2019raman
}.
Moreover, Joule heating can cause less trivial effects in measurements.
For example, current-induced diamagnetism reported by our group \cite{
	sow2017current,
	sow2019situ,
	sow2020retractionscience,
	sow2020retractionprl
},
was then revealed to be caused by localised heating of the sample holder \cite{
	mattoni2020diamagnetic
}.
Also, there are cases in which the occurrence of MITs in other materials has been linked to local heating effects \cite{
	fursina2009origin,
	zimmers2013role,
	shukla2014electrically,
	mattoni2018light
}.
Due to these considerations, a local temperature probe allowing direct imaging of \ce{Ca2RuO4} is strongly desirable to uncover current-induced changes in the MIT.

Here, we study the current-driven MIT in \ce{Ca2RuO4} by means of infrared (IR) thermal imaging.
Our technique allows us to produce spatial maps of the local sample temperature, as well as to image the coexistence of metallic and insulating regions by making use of their different reflectivity.
We uncover that the metallic phase forms and extends as a function of continuous electrical current, and that a sample area becomes metallic or insulating whenever its local temperature crosses the characteristic MIT temperatures.
Thus, within our experimental resolution, the value of these transition temperatures remains unchanged by the flowing current.
Our imaging also allows us to reveal the formation of macroscopic cracks and defects during repeated cycles of current-driven MIT, which we find to strongly affect the MIT spatial evolution.
These findings uncover the important role of local sample temperature in triggering phase transitions, in particular under the presence of direct current.

\section{Results and discussion}
\subsection{Thermal imaging setup}

\begin{figure}[tb]
\includegraphics[page=1,width=80mm]{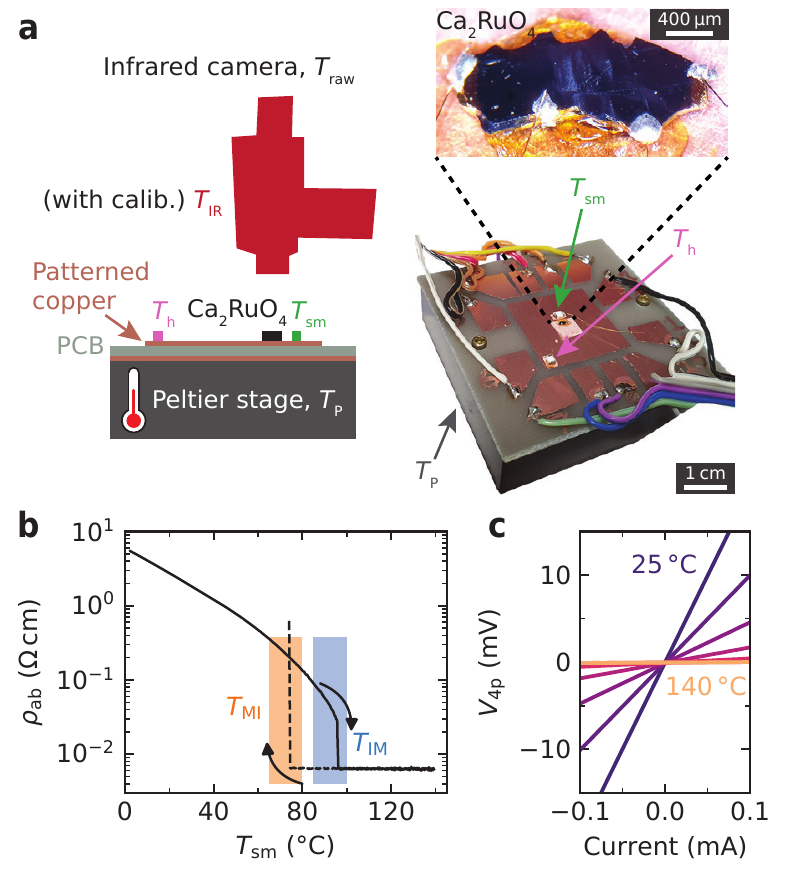}

\subfloat{\label{fig:IR_camera}}
\subfloat{\label{fig:RvsT}}
\subfloat{\label{fig:IVs}}

\caption{\textbf{Setup for thermal imaging.}
		\protect\subref{fig:IR_camera} Schematics and photographs of the thermal imaging setup, the sample holder, and a typical \ce{Ca2RuO4} single crystal (Sample \#1) with electrical leads in four-probe configuration.
		\protect\subref{fig:RvsT} Resistivity of \ce{Ca2RuO4} as a function of $\Tsm$ controlled by varying the Peltier stage.
		The shaded regions show the range of temperatures where the insulator-to-metal ($\Tim$, blue) and metal-to-insulator ($\Tmi$, orange) transitions are observed in our samples.
		Upon cooling (dashed line), a hysteresis is observed and the sample often breaks at $\Tmi$.
		\protect\subref{fig:IVs} Four-probe voltage--current characteristics in the low-current regime.
		A linear trend is observed at all temperatures, both in the insulating and in the metallic phases.
}

\label{fig:Setup}

\end{figure}

For the experiment, we used the thermal imaging setup in \cref{fig:IR_camera}.
An IR camera (Avio, InfRec R500, spatial resolution \SI{21}{\micro\metre}, spectral range \SIrange{8}{14}{\micro\metre}) allows to acquires optical maps of the raw IR reading $\Tr$ that can be converted to the sample IR temperature $T_\mathrm{IR}$ upon calibration.
A Peltier stage (Ampère, UT4070 with UTC-200A temperature controller) allows to control the temperature of the thermal bath ($\Tp$) to which the sample is anchored.
The sample holder consists of an FR-4 glass epoxy plate (Sunhayato, 31R, thickness \SI{1.6}{\milli\metre}) covered on each face by a thin copper layer (\SI{35}{\micro\metre}), which is patterned to provide contact pads on the top face.
Two Pt1000 sensors (Heraeus, SMD 805-1000-B) are glued on the top copper layer with GE varnish in order to measure the temperature at \SI{1}{\milli\metre} (sample monitor, $\Tsm$) and at \SI{15}{\milli\metre} (holder, $\Th$) from the sample.
The copper layer enhances the thermal coupling between the sensors and the sample.
We use high-purity \ce{Ca2RuO4} single crystals of typical dimension between \SI{0.6x0.5x0.1}{} and \SI{2x1x0.3}{\milli\metre}, with an exposed $a_\mathrm{o}b_\mathrm{o}$ surface in orthorhombic notation (\cref{fig:Characterisation}).
The sample is glued to the holder by using GE varnish with a layer of cigarette paper in between for electrical insulation.
This configuration provides a strong thermal coupling, maximising the cooling of the sample.
To obtain low-resistance ohmic contacts, a thin layer of \ce{Au} is sputtered at a \ang{45} angle on the edges of the sample, and then connected to gold wires (thickness \SI{18}{\micro\metre}) by using silver epoxy (Epotek, H20E).
We measure a contact resistance smaller than \SI{100}{\ohm} at room temperature (\cref{fig:ContactResistance}).
The electrical measurements are performed using a variable current source (Keysight, B2912A) and multimeters to measure the sample current and voltage (Keithley, 2000).

In \cref{fig:RvsT}, we show the typical resistivity of a \ce{Ca2RuO4} sample measured in a four-probe configuration with constant current $I=\SI{10}{\micro\ampere}$ (current density $j\sim\SI{0.01}{\ampere\per\square\centi\metre}$).
The sample temperature is measured by $\Tsm$ while varying the Peltier-stage temperature at \SI{1}{\kelvin\per\minute}.
Upon warming, the resistivity shows the expected insulating trend, with a sharp drop at $\Tim$, where the sample becomes metallic.
Upon cooling, the resistivity presents a clear hysteresis, consistent with previous results on \ce{Ca2RuO4} \cite{
	friedt2001structural,
	steffens2005high
},
and typical of first order phase transitions \cite{
	morin1959oxides,
	mattoni2016striped
}.
At $\Tmi$, the sample turns back insulating and we often lose the electrical contacts because the crystal shatters (vertical dashed line).
The values of in-plane resistivity for the insulating ($\rho_{ab}\sim\SI{1}{\ohm\centi\metre}$ at room temperature) and metallic ($\rho_{ab}\sim\SI{e-2}{\ohm\centi\metre}$) phases indicate high-purity \ce{Ca2RuO4} single crystals \cite{
	alexander1999destruction,
	cao2000ground,
	fobes2012metal
}.
As shown in \cref{fig:IVs}, the $I$--$V$ characteristics up to \SI{\pm 100}{\micro\ampere} (current density \SI{\pm0.1}{\ampere\per\square\centi\metre}) are linear, indicating that the sample is in an ohmic regime in the explored current range, both in the insulating and metallic phases.

The MIT of \ce{Ca2RuO4} is rather abrupt and is known to involve a large change of the lattice parameters, with \SI{+1}{\percent} change of $c_\mathrm{o}$, and up to \SI{-1}{\percent} change of $a_\mathrm{o}$ and $b_\mathrm{o}$ at $\Tim$ \cite{
	friedt2001structural
}.
The \ce{Ca2RuO4} samples often cleave upon warming and, more frequently, shatter in several pieces upon cooling \cite{
	alexander1999destruction
}.
This is naturally understood as due to the material fragility to expansion, which occurs in the out-of-plane direction upon warming and in-plane upon cooling.
Consequently, we need to use multiple samples for our study (\cref{fig:Samples}), each one having its own $\Tim$ and $\Tmi$, whose values vary in the ranges indicated by the shaded regions of \cref{fig:RvsT} and throughout this manuscript ($\Tim = \text{\SIrange{85}{100}{\celsius}}$ in blue, $\Tmi = \text{\SIrange{65}{80}{\celsius}}$ in orange).
These sample-to-sample variations can be due to slight differences in chemical stoichiometry, in the amount of micro-cracks, or in strain possibly caused by the glueing on the holder.

\subsection{Temperature-driven MIT}
\begin{figure}[tbh]
\includegraphics[page=2,width=86mm]{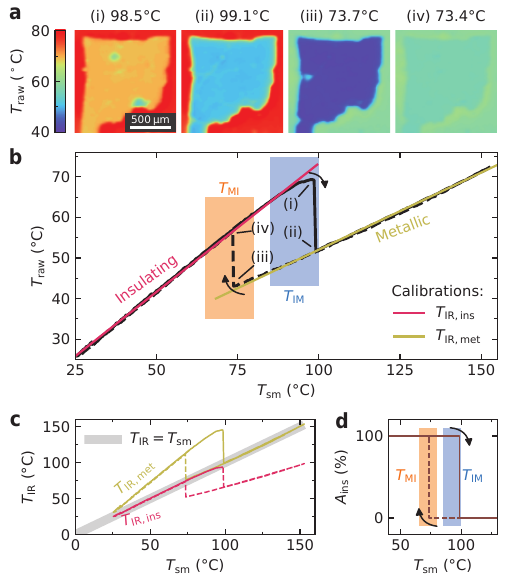}

\subfloat{\label{fig:TIR_Images}}
\subfloat{\label{fig:TIR_Calibration}}
\subfloat{\label{fig:TIR_Calibrated}}
\subfloat{\label{fig:ThermalMIT_Area}}

\caption{\textbf{Thermal imaging of the temperature-driven MIT in \ce{Ca2RuO4} with $I=0$.}
		\protect\subref{fig:TIR_Images} Thermal images taken at different values of $\Tsm$ as indicated (Sample \#2, pristine state, no electrical leads).
		Images (i) and (ii) show the insulator-to-metal transition upon warming, while images (iii) and (iv) the metal-to-insulator upon cooling.
		\protect\subref{fig:TIR_Calibration} Raw reading of the IR camera $\Tr$ averaged over the sample surface as a function of $\Tsm$.
		The jumps in $\Tr$ are due to the abrupt change in \ce{Ca2RuO4} reflectivity at the MIT.
		Two linear regressions are used to separately calibrate $T_\mathrm{IR}$ for the insulating (red) and metallic (gold) phases.
		\protect\subref{fig:TIR_Calibrated} IR temperature calculated from $\Tr$ by using the two calibrations obtained in \protect\subref{fig:TIR_Calibration}.
		Overlap with the grey line shows that the calibration $\Tins$ ($\Tmet$) is valid only when the sample is in the insulating (metallic) phase.
		\protect\subref{fig:ThermalMIT_Area} Percentage of insulating area $\Ai$ as a function of temperature estimated by making use of the emissivity change.
		In this temperature-driven MIT, the whole sample changes phase at the same time.
}

\label{fig:ThermalMIT}

\end{figure}

We now perform thermal imaging on \ce{Ca2RuO4} during a warming-and-cooling ramp without any supplied current ($I=0$).
The representative thermal images in \cref{fig:TIR_Images} show that the raw IR reading $\Tr$ is rather homogeneous over the sample surface.
In order to measure the appropriate IR temperature, $\Tr$ needs to be calibrated according to the emissivity of \ce{Ca2RuO4}.
To perform this calibration, we exploit the fact that in these conditions the sample is in thermal equilibrium with $\Tsm$.
We thus extract in \cref{fig:TIR_Calibration} the dependence of $\Tr$, averaged over the sample surface, on $\Tsm$.
Upon warming, $\Tr$ shows a sharp drop at $\Tim$ when the sample becomes metallic.
This drop is due to the decreased thermal emissivity of \ce{Ca2RuO4} in the metallic phase, in accordance with its increased reflectivity \cite{
	zhang2019nano
}.
Upon cooling, $\Tr$ rises sharply at $\Tmi$ when the sample turns insulating, and it goes back to the previous curve.

The observed experimental behaviour prompts for two different emissivity calibrations for the two phases.
These are obtained by performing a linear regression for the insulating ($\Tins$, red line) and for the metallic ($\Tmet$, gold line) phases separately.
We note that the regressions have some deviation from the experimental data, especially in the insulating phase close to $\Tim$.
Furthermore, there are two additional sources of error that affect our calibration:
the thermal contribution due to areas surrounding the sample,
and sample-to-sample variations (\cref{fig:EmissivityCalibration}).
For these reasons, we estimate an uncertainty of about \SI{\pm10}{\celsius} on the absolute temperature value provided by the calibrations.
Since the same calibrations are used throughout this work, this error is of systematic character and it does not affect our findings.
We comment that a smaller error on $T_\mathrm{IR}$ has been previously achieved by coating the sample with a blackbody insulating paste \cite{
	okazaki2013current
}.
Nevertheless, our experimental configuration has the important advantage of allowing direct imaging of metallic and insulating areas, as discussed in the following.

We use the obtained calibrations to convert $\Tr$ into the appropriate IR temperature for the insulating and metallic phases in \cref{fig:TIR_Calibrated}.
As a reference, we also plot $T_\mathrm{IR} = \Tsm$ (thick grey line), which is the result expected from an ideal calibration.
\Cref{fig:TIR_Calibrated} shows that $\Tins$ has a good overlap with the grey line as long as the sample is in the insulating phase.
In the metallic state, instead, $\Tins$ is significantly lower than the effective sample temperature, whereas $\Tmet$ overlaps the grey line.

The large change in emissivity at the MIT allow us to map the local distribution of metallic and insulating phases.
Areas turning from insulating to metallic, in fact, suddenly appear darker in our images.
For the case of the temperature-driven MIT presented here, the whole sample changes phase at the same time, as shown by the inset images of \cref{fig:TIR_Calibration}.
This is a direct consequence of the spatially homogeneous temperature distribution over the sample, which is brought above $\Tim$ or below $\Tmi$ all at the same time.
In addition, the single-crystal nature of the sample contributes to a simultaneous phase transition in the entire volume.
For these reasons, the percentage of insulating area $\Ai$ as a function of temperature has a step-like trend (\cref{fig:ThermalMIT_Area}).
As shown by the overlap with the rectangular shaded areas, $\Ai$ has a hysteresis compatible with the transport data of \cref{fig:RvsT}.

\begin{figure*}[tb]
\includegraphics[page=3,width=176mm]{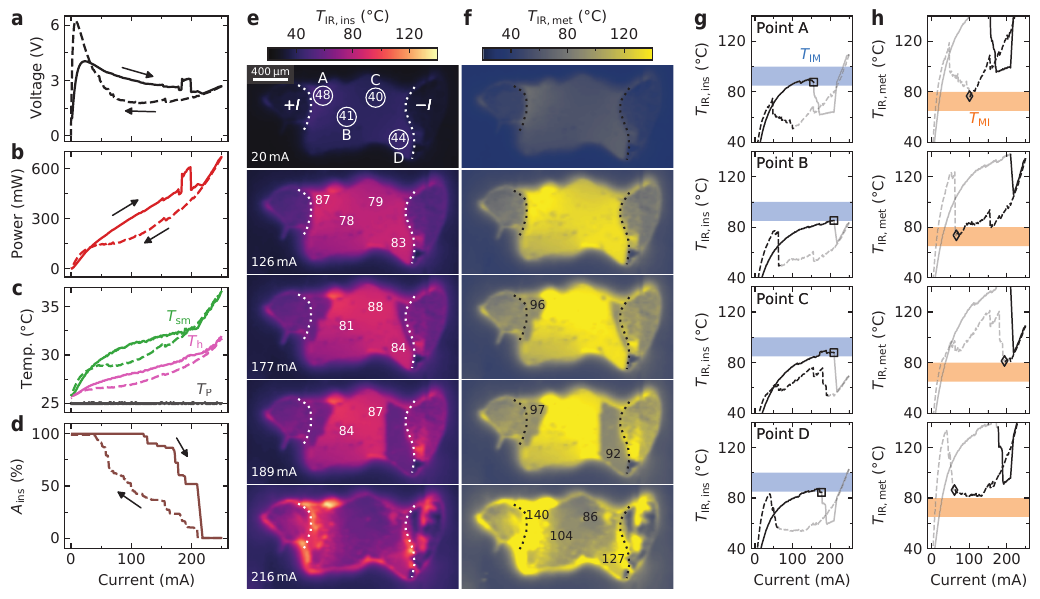}

\subfloat{\label{fig:CurrentMIT_Voltage}}
\subfloat{\label{fig:CurrentMIT_Power}}
\subfloat{\label{fig:CurrentMIT_Temperature}}
\subfloat{\label{fig:CurrentMIT_Area}}
\subfloat{\label{fig:CurrentMIT_IR_ins}}
\subfloat{\label{fig:CurrentMIT_IR_met}}
\subfloat{\label{fig:CurrentMIT_points_ins}}
\subfloat{\label{fig:CurrentMIT_points_met}}

\caption{\textbf{Current-driven MIT in \ce{Ca2RuO4}.}
		\protect\subref{fig:CurrentMIT_Voltage} Voltage--current characteristics measured in a two probe configuration and
		\protect\subref{fig:CurrentMIT_Power} corresponding electrical power (Sample \#3).
		\protect\subref{fig:CurrentMIT_Temperature} Temperature measured by the sample-monitor and holder sensors at a fixed $\Tp=\SI{25}{\celsius}$.
		\protect\subref{fig:CurrentMIT_Area} Percentage of insulating area extracted from the thermal images.
		\protect\subref{fig:CurrentMIT_IR_ins} Thermal images with insulating and
		\protect\subref{fig:CurrentMIT_IR_met} metallic calibration.
		The temperature value of points A to D is indicated by a number in the image corresponding to the appropriate state of the point.
		The current electrodes are indicated by $\pm I$ and the dotted lines.
		\protect\subref{fig:CurrentMIT_points_ins} Current dependence of the IR temperature obtained with the insulating and
		\protect\subref{fig:CurrentMIT_points_met} metallic calibration for the points indicated in 		\protect\subref{fig:CurrentMIT_IR_ins}.
}

\label{fig:CurrentMIT}

\end{figure*}

\subsection{Current-driven MIT}

We now investigate how the electric current can drive the MIT in \ce{Ca2RuO4}.
For this measurement, we fix the Peltier-stage temperature to \SI{25}{\celsius} in air environment and drive the sample with a continuous current source.
We use a two-probe configuration in order to measure the voltage and power supplied to the whole sample.
As shown in \cref{fig:CurrentMIT_Voltage}, the voltage--current characteristics has an initial linear trend, peaks at about \SI{20}{\milli\ampere}, and then evolves into a negative differential resistance.
We observed a similar behaviour also when driving the sample with a voltage source or a voltage divider (\cref{fig:DifferentDrives}).
This voltage--current characteristics is consistent with the one observed in several previous experiments on \ce{Ca2RuO4} under current flow \cite{
	nakamura2013electric,
	okazaki2013current,
	shukla2016electrically,
	zhang2019nano,
	cirillo2019emergence,
	fursich2019raman
}.
With increasing current, the supplied electrical power increases (\cref{fig:CurrentMIT_Power}).
Consequently, both $\Tsm$ and $\Th$ detect a small temperature increase of a few degrees (\cref{fig:CurrentMIT_Temperature}).
At the same time, however, the actual sample temperature increases to a much larger extent: $\Tins$ shows a raise of several tens of \SI{}{\celsius} as measured by the IR camera in \cref{fig:CurrentMIT_IR_ins}.
This indicates that the temperature sensors placed on the sample holder largely underestimate the actual temperature of \ce{Ca2RuO4} which is heated up by the direct current, as also reported by a recent study by K. F\"ursich \textit{et al.} \cite{
	fursich2019raman
}.
This observation remarks the need for monitoring the local sample temperature as allowed by our imaging technique.

The temperature distribution over the sample surface is approximately homogeneous, apart from slightly hotter regions near the contacts, probably due to the higher current density dictated by the sample geometry.
The low-resistance gold pads used for the electrical contacts guarantee a small power dissipation due to contact resistance (\cref{fig:ContactResistance}).
In another experimental configuration where the sample has a weaker thermal coupling with the cooling stage, we observe a Peltier effect at metal--\ce{Ca2RuO4} interfaces, which produces a higher temperature at the negative current electrode (\cref{fig:PeltierHeating}).
This Peltier effect and its sign are consistent with previous experimental and theoretical reports both on \ce{Ca2RuO4} \cite{
	zhang2019nano,
	chiriaco2020polarity
} and \ce{VO2} \cite{
	favaloro2014direct
}.
In the data presented here, Peltier effects are negligible due to the strong thermal coupling between the sample and the holder.

Upon increasing current, the metallic phase nucleates simultaneously at both current electrodes, from which it expands starting from the hotter regions close to the contacts (see the darker areas in the 3rd panel from the top in \cref{fig:CurrentMIT_IR_ins}).
We comment that in similar conditions nanoscale stripes at the phase front were reported by Zhang \textit{et al.} \cite{
	zhang2019nano
}, but we cannot resolve them with the present technique.
The change in thermal emissivity gives us the unique possibility of extracting from the images $\Ai$ (\cref{fig:CurrentMIT_Area}), which progressively decreases with increasing current as the sample becomes more metallic.
The appropriate temperature of metallic areas is provided by the calibration $\Tmet$ shown in \cref{fig:CurrentMIT_IR_met}.
Once the metallic phase appears, the resistivity is locally reduced resulting in a lower local dissipation of electrical power.
This effect is expected to produce a local cooling, but in reality it leads to complex rearrangements of the effective current path, sometimes causing a local switch back to the insulating phase, or to repeated insulating/metallic switches.
In most cases, however, the MIT hysteresis prevents the newly-formed metallic areas from going back to the insulating state.
Due to the \SI{\pm10}{\celsius} uncertainty in our temperature calibrations, we cannot experimentally assess whether regions turning metallic experience a local cooling.

For $I>\SI{215}{\milli\ampere}$ the entire sample is in the metallic state.
Upon decreasing current, the sample cools down and progressively goes back to the insulating state.
At this point, the curves in \cref{fig:CurrentMIT_Voltage,fig:CurrentMIT_Power,fig:CurrentMIT_Temperature,fig:CurrentMIT_Area} show a hysteresis which can be due to various effects.
Firstly, there is the intrinsic hysteresis of the MIT as discussed in \cref{fig:RvsT}.
Secondly, when the current is decreased from the maximum value the sample is in the less-resistive metallic state, the voltage, the power, and the sensor temperatures also have lower values.
Thirdly, the current-induced MIT leads to the irreversible formation of cracks and defects, as will be discussed in the following subsection.
The presence of this irreversible effect is also evidenced by the fact that once the sample goes back to the insulating state at low current, the peak in the $V$--$I$ curve (\cref{fig:CurrentMIT_Voltage}) occurs at a higher voltage value and the sample has a higher resistivity.

Our imaging technique allows us to extract the local temperature of some representative points (A--D in \cref{fig:CurrentMIT_IR_ins}) as a function of applied current.
We report in \cref{fig:CurrentMIT_IR_ins} and \cref{fig:CurrentMIT_IR_met} the local temperature of these points using the insulating and metallic calibrations, respectively.
We use a solid line for the increasing-current ramp and a dashed line for the decreasing current.
Data with the calibration matching the appropriate phase is in black colour, while the rest is in shaded grey.
The curves in \cref{fig:CurrentMIT_IR_ins} show the local temperature increase due to the increasing current up to the point where the insulator-to-metal transition occurs (open squares).
The opposite phase transition (metal-to-insulator) can be observed in \cref{fig:CurrentMIT_IR_ins} upon decreasing current, as indicated by the open diamonds.
We note that the square and diamond markers lie, within our experimental resolution, within the ranges of $\Tim$ and $\Tmi$, respectively.
This indicates that the current-driven MIT is triggered whenever the local temperature of an area goes above $\Tim$ or below $\Tmi$.
The experimental setup used in the present work does not allow us to detect additional non-thermal effects induced by the electrical current, which may have caused changes in $\Tim$ and $\Tmi$ smaller than our uncertainty of \SI{\pm10}{\celsius}.

\subsection{Current-driven MIT in the presence of cracks and defects}
\begin{figure}[tb]
\includegraphics[page=4,width=86mm]{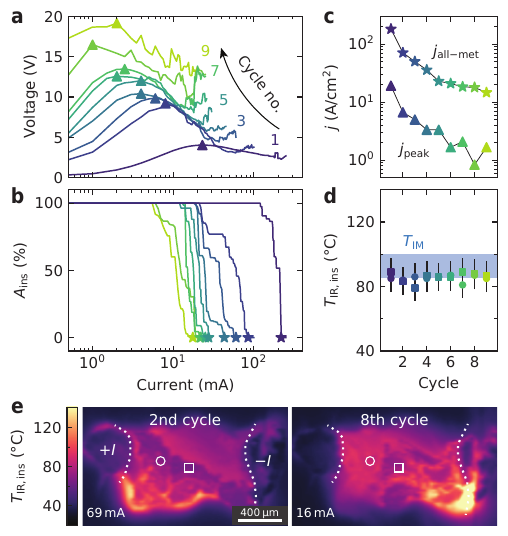}

\subfloat{\label{fig:Fatigue_V}}
\subfloat{\label{fig:Fatigue_A}}
\subfloat{\label{fig:Fatigue_j}}
\subfloat{\label{fig:Fatigue_T}}
\subfloat{\label{fig:Fatigue_imgs}}

\caption{\textbf{Multiple cycles of current-driven MIT.}
		\protect\subref{fig:Fatigue_V} Current dependence of the two-probe voltage and
		\protect\subref{fig:Fatigue_A} percentage of insulating area for the different increasing-current sweeps (Sample \#3).
		\protect\subref{fig:Fatigue_j} Current density at the voltage peak (triangles) and when the sample becomes all metallic (stars) as a function of cycle number.
		\protect\subref{fig:Fatigue_T} IR temperature at which the regions indicated by the circle and the square in \protect\subref{fig:Fatigue_imgs} turn metallic as a function of cycle number.
		\protect\subref{fig:Fatigue_imgs} Representative thermal images at different current cycles during the increasing ramp.
		The presence of a hot spot in the bottom left (right) corner during the 2nd (8th) current cycle drives the formation of the metallic phase from the left (right) corner.
}

\label{fig:Fatigue}

\end{figure}

To study how the formation of cracks and defects affects the current-driven MIT, we perform multiple current cycles that bring the sample to the metallic state and then back to the insulating state.
The thermal bath is fixed at $\Tp=\SI{25}{\celsius}$.
The current is increased up to a few \SI{}{\milli\ampere} larger than what required to switch the entire sample to the metallic phase, and then decreased back to zero (rates of \SI{0.5}{\milli\ampere\per\second}).
We show in \cref{fig:Fatigue_V,fig:Fatigue_A} the two-probe $V$--$I$ characteristics and value of $\Ai$ measured during nine consecutive cycles.
At each cycle, the voltage peak increases and shifts to lower currents (triangles in \cref{fig:Fatigue_V}) and to lower corresponding current densities ($\jp$ in \cref{fig:Fatigue_j}).
Concomitantly, the entire sample turns metallic at a lower current (stars in \cref{fig:Fatigue_A}) and current density ($\ja$ in \cref{fig:Fatigue_j}).
This occurs because a higher power is dissipated in the sample due to its increased resistance.
The drop of more than one order of magnitude in $\ja$ indicates large and irreversible changes in the sample properties, consistent with the previous observation of crack formation at similar current values \cite{
	sakaki2013electric
}.

The formation of cracks and defects strongly affects the temperature distribution over the sample surface, leading to the formation of hot spots responsible for higher power dissipation, as can be seen in the images of \cref{fig:Fatigue_imgs}.
It is clear that, despite the comparable values of $\Ai$, the presence of hot spots in the lower left (right) corner during the 2nd (8th) cycle drives a preferential metallic phase formation from the left (right) electrode.
We select two representative regions, indicated by the circle and square in \cref{fig:Fatigue_imgs}, and report in \cref{fig:Fatigue_T} the values of $\Tins$ at which they become metallic at each current cycle.
The values of $\Tins$ are rather constant and in good agreement with the temperature range of $\Tim$ (shaded blue area in \cref{fig:Fatigue_T}), even if the insulator-to-metal transition in these regions occurs for very different current densities at each cycle.
This data is a further indication that, in our experimental conditions and irrespective of the sample state, the main trigger for the current-induced MIT is the local sample temperature.

\section{Conclusions}
To conclude, we were able to observe the current-induced formation of metallic and insulating regions in \ce{Ca2RuO4} by means of thermal IR imaging.
Our data shows that while the insulator-to-metal transition can occur over a broad range of current densities, it is always triggered whenever the local sample temperature exceeds $\Tmi$.
Within our resolution, the local $\Tmi$ is the same regardless if the MIT is triggered by temperature, by current, or by current in the presence of macroscopic cracks and defects.
In our experimental conditions we find that direct heating of the sample due to the Joule effect is much larger than what is detected by thermometers that monitor the sample holder temperature.
These findings uncover an important balance of local heating, phase coexistence, and microscale inhomogeneity in \ce{Ca2RuO4} that is of utmost importance when performing measurements with applied current.
These effects can potentially have a large influence also on macroscopic measurements, that may average the contribution of coexisting metallic and insulating regions.
The present work sets an important basis to perform future experiments to uncover current-induced non-equilibrium steady-state effects.

\section*{Acknowledgements}
The authors thank
G. Chiriac\`{o},
K. F\"{u}rsich,
M. Liu,
N. Manca,
A. Millis,
H. Narita,
R. Okazaki,
D. Ootsuki,
I. Terasaki,
T. Yoshida, and
A. Vecchione
for fruitful discussions.
This work was supported by JSPS Grant-in-Aids KAKENHI Nos. JP26247060, JP15H05852, JP15K21717, and JP17H06136, as well as by JSPS Core-to-Core program.
G.M. acknowledges support from the Dutch Research Council (NWO) through a Rubicon grant number 019.183EN.031.

\bibliography{%
	Biblio_NonEquilibrium,
	Biblio_QMaterials,
	Biblio_Ruthenates,
}

\ifsuppinpaper

\ifsuppinpaper

	\onecolumngrid
	\appendix
	\newpage
	\noindent\rule{1\columnwidth}{1pt}
	\section*{\huge \texttt{Supporting Information}}
	\noindent\rule{1\columnwidth}{1pt}

\else

	\documentclass[aps,nobibnotes,superscriptaddress]{revtex4-1}



\usepackage{graphicx} 	
\usepackage{dcolumn} 	
\usepackage{bm}   		
\usepackage{amssymb}   	
\usepackage[separate-uncertainty=true,range-units=single,range-phrase=--,retain-explicit-plus=true]{siunitx} 
\usepackage[svgnames]{xcolor}	
\usepackage[version=3]{mhchem}	

\hyphenation{ALPGEN}
\hyphenation{EVTGEN}
\hyphenation{PYTHIA}

\usepackage{xr} 
\usepackage[colorlinks,
citecolor=Green,urlcolor=RoyalBlue]{hyperref}
\usepackage[caption=false]{subfig}
\usepackage[nameinlink,capitalise]{cleveref}
\usepackage{lipsum}

\newcommand{\ch}{$C_{\ce{H2}}$}
\DeclareSIUnit\unitcell{uc}

\graphicspath{{figs/}}

	\def\Tim{T_\mathrm{IM}}
	\def\Tmi{T_\mathrm{MI}}
	\def\Tp{T_\mathrm{P}}
	\def\Th{T_\mathrm{h}}
	\def\Tsm{T_\mathrm{sm}}
	\def\Tr{T_\mathrm{raw}}
	\def\Tins{T_\mathrm{IR,\,ins}}
	\def\Tmet{T_\mathrm{IR,\,met}}
	\def\Ai{A_\mathrm{ins}}
	\def\jp{j_\mathrm{peak}}
	\def\ja{j_\mathrm{all\text{-}met}}
	\externaldocument{CRO_IRimaging}
	\begin{document}
	\title{\texttt{Supporting Information}\\
		Role of local temperature in the current-driven metal--insulator transition of \ce{Ca2RuO4}
	}
	\author{Giordano Mattoni}
\email{mattoni@scphys.kyoto-u.ac.jp}
\affiliation{Department of Physics, Graduate School of Science, Kyoto University, Kyoto 606-8502, Japan}

\author{Shingo Yonezawa}
\affiliation{Department of Physics, Graduate School of Science, Kyoto University, Kyoto 606-8502, Japan}

\author{Fumihiko Nakamura}
\affiliation{Department of Education and Creation Engineering, Kurume Institute of Technology, Fukuoka 830-0052, Japan}

\author{Yoshiteru Maeno}
\affiliation{Department of Physics, Graduate School of Science, Kyoto University, Kyoto 606-8502, Japan}

	\maketitle
	
\fi

\renewcommand\thefigure{S\arabic{figure}}    
\setcounter{figure}{0}
\renewcommand\thetable{S\arabic{table}}    
\setcounter{table}{0}
\renewcommand\theequation{S\arabic{equation}}    
\setcounter{equation}{0}

\begin{figure*}[h]
	\includegraphics[page=5,width=130mm]{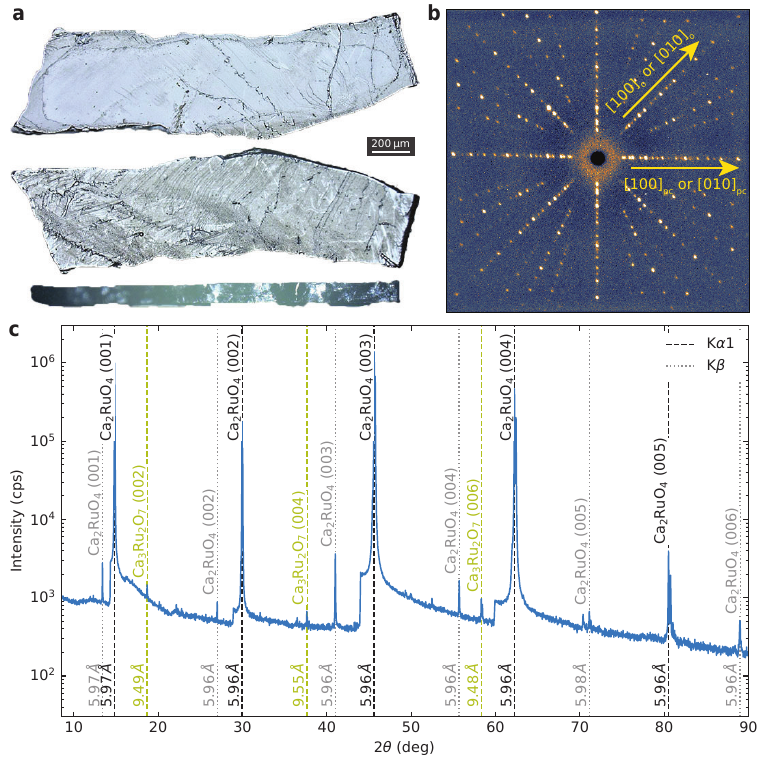}
	
	\subfloat{\label{fig:Characterisation_photo}}
	\subfloat{\label{fig:Characterisation_Laue}}
	\subfloat{\label{fig:Characterisation_XRD}}
	
	\caption{\textbf{Structural characterisation of a typical \ce{Ca2RuO4} sample.}
		\protect\subref{fig:Characterisation_photo} Optical image of the front, the back, and the side surfaces (Sample \#1, CR29C1a).
		The front surface has been mechanically cleaned with acetone and a cotton bud, while the bottom surface is still covered by an opaque layer, possibly due to the natural formation of surface hydroxides.
		\protect\subref{fig:Characterisation_Laue} Laue photograph of the front surface taken with the same sample orientation shown in \protect\subref{fig:Characterisation_photo}.
		The pseudocubic (pc) and orthorhombic (o) lattice directions are indicated by the arrows.
		All samples investigated in this work have an exposed $c$ plane, and the current is sourced in the $a_\mathrm{o}b_\mathrm{o}$ plane.
		\protect\subref{fig:Characterisation_XRD} X-ray diffraction in $\theta$--$2\theta$ configuration showing intense peaks of the high-purity \ce{Ca2RuO4} phase taken with a standard \ce{Cu} X-ray source.
		Due to the absence of an X-ray monochromator, we observe both the K$\alpha$1, K$\beta$, and the sharp edge of the notch filter at wavelengths slightly smaller than K$\alpha$1.
		The only impurity detected is a minor amount of the \ce{Ca3Ru2O7} phase, with diffraction peaks more than 2 orders of magnitude less intense.
	}
	
	\label{fig:Characterisation}
	
\end{figure*}

\begin{figure*}[h]
	\includegraphics[page=6,width=165mm]{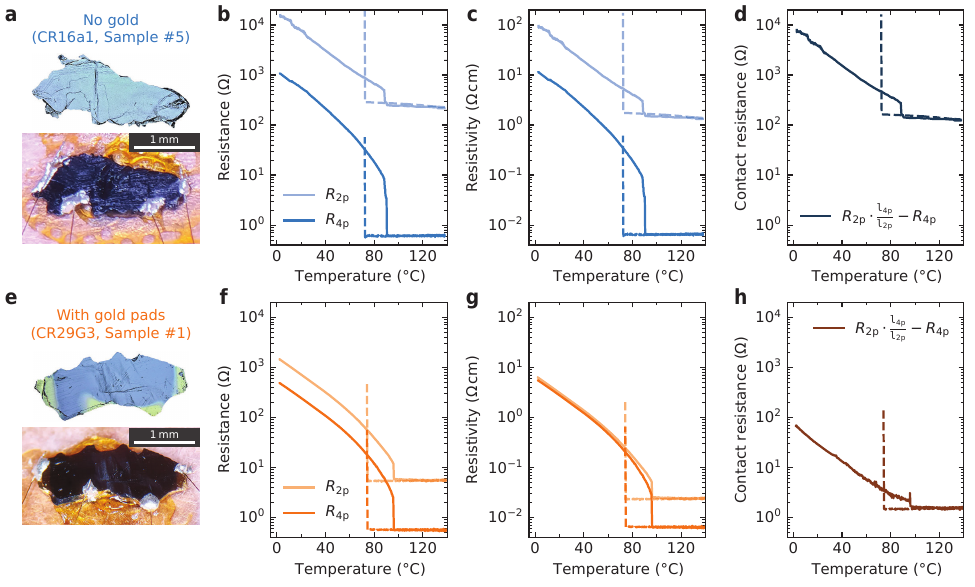}
	
	\subfloat{\label{fig:ContactResistance_NoAu}}
	\subfloat{\label{fig:ContactResistance_NoAu_R}}
	\subfloat{\label{fig:ContactResistance_NoAu_rho}}
	\subfloat{\label{fig:ContactResistance_NoAu_delta}}
	\subfloat{\label{fig:ContactResistance_WithAu}}
	\subfloat{\label{fig:ContactResistance_WithAu_R}}
	\subfloat{\label{fig:ContactResistance_WithAu_rho}}
	\subfloat{\label{fig:ContactResistance_WithAu_delta}}	
	
	\caption{\textbf{Contact resistance and gold sputtering.}
		\protect\subref{fig:ContactResistance_NoAu} Optical image of a \ce{Ca2RuO4} single crystal (Sample \#4, CR16a1) before and after the electrodes fabrication.
		In this case, \ce{Ag} epoxy is directly placed on the sample surface and edges to provide electrical contact with the wires.
		\protect\subref{fig:ContactResistance_NoAu_R} Resistance ($R_\mathrm{2p}$, $R_\mathrm{4p}$) and
		\protect\subref{fig:ContactResistance_NoAu_rho} resistivity ($\rho_\mathrm{2p}$, $\rho_\mathrm{4p}$) measured in two-probe configuration (shaded curve) using the outer electrodes, and in four-probe configuration (solid curve) using the inner electrodes.
		\protect\subref{fig:ContactResistance_NoAu_delta} Contact resistance calculated by subtracting $R_\mathrm{4p}$ from $R_\mathrm{2p}$ rescaled by the different channel length $l_\mathrm{4p}/l_\mathrm{2p}$.
		\protect\subref{fig:ContactResistance_WithAu} Optical image of a \ce{Ca2RuO4} single crystal (Sample \#5, CR29G3) after sputtering of gold electrodes and subsequent contacting with \ce{Ag} epoxy.
		The gold sputtering (\ce{Ar} environment, \SI{8}{\minute} at \SI{5}{\milli\ampere} ionic current) is performed at a 45-degree angle in order to cover both the top and the side surfaces of the sample.
		A gold foil (thickness \SI{10}{\micro\metre}) is used as a shadow mask to define the contacts area.
		\protect\subref{fig:ContactResistance_WithAu_R} Resistance,
		\protect\subref{fig:ContactResistance_WithAu_rho} resistivity, and
		\protect\subref{fig:ContactResistance_WithAu_delta} contact resistance for the sample with gold pads.
		The four-probe resistivity of the two samples is comparable, and it is consistent with the values reported in literature.
		We observe a reduction of about two orders of magnitude in contact resistance in the presence of gold contact pads.
		We note how the contact resistance is rather negligible compared to the sample resistance in the insulating phase, so that power dissipation in the contacts provides a small contribution to the Joule heating.
	}
	
	\label{fig:ContactResistance}
	
\end{figure*}

\begin{figure*}[h]
	\includegraphics[page=7,width=120mm]{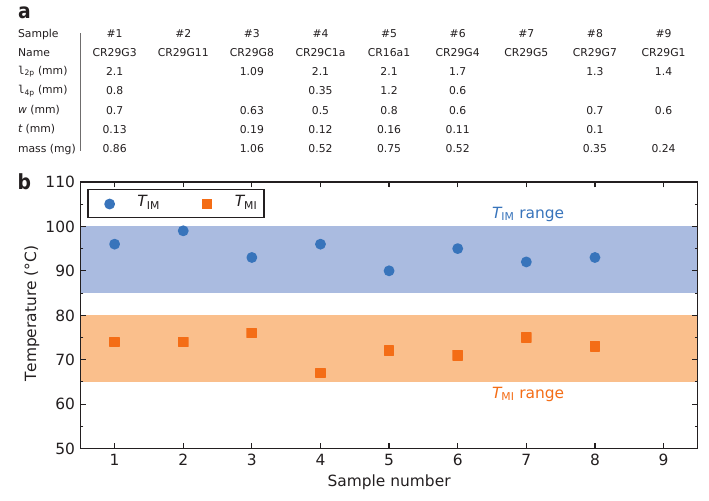}
	
	\subfloat{\label{fig:Samples_List}}
	\subfloat{\label{fig:Samples_TMIT}}
	
	\caption{\textbf{Properties of the main samples used in this work.}
	\protect\subref{fig:Samples_List} Table of samples including the distance between the current ($l_\mathrm{2p}$) and voltage ($l_\mathrm{4p}$) probes, the sample width ($w$), the thickness ($t$), and the mass measured prior to the fabrication of the contacts.
	\protect\subref{fig:Samples_TMIT} Insulator-to-metal ($\Tim$, blue circles) and metal-to-insulator ($\Tmi$, orange squares) transition temperatures measured upon thermally cycling the pristine samples.
	The shaded regions indicate the following ranges of transition temperature: \SIrange{85}{100}{\celsius} for $\Tim$, and \SIrange{65}{80}{\celsius} for $\Tmi$.
	}
	
	\label{fig:Samples}
	
\end{figure*}

\begin{figure*}[h]
	\includegraphics[page=8,width=150mm]{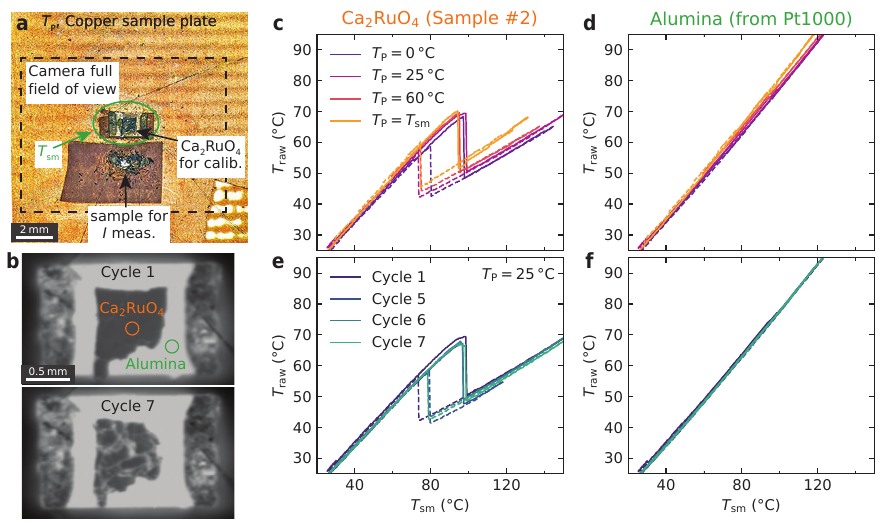}
	
	\subfloat{\label{fig:EmissivityCalibration_Holder}}
	\subfloat{\label{fig:EmissivityCalibration_Photos}}
	\subfloat{\label{fig:EmissivityCalibration_SampleT}}
	\subfloat{\label{fig:EmissivityCalibration_AluminaT}}
	\subfloat{\label{fig:EmissivityCalibration_SampleCycle}}
	\subfloat{\label{fig:EmissivityCalibration_AluminaCycle}}

	\caption{\textbf{Details on the calibration of the infrared thermal emissivity.}
	\protect\subref{fig:EmissivityCalibration_Holder} Experimental configuration used to calibrate the thermal emissivity.
	The copper sample plate is directly attached to the Peltier controller which is set to a temperature $\Tp$.
	The full field of view of the thermal camera is indicated by the black dashed line: this is the portion of space that reaches the camera optics and imaging sensor, determining the background of the measured signal.
	A \ce{Ca2RuO4} sample for calibration is glued on top of the sample-monitor Pt1000 sensor (green circle, $\Tsm$) with GE varnish.
	Two wires are used to source a current to the sensor and to measure $\Tsm$.
	Another (broken) sample, previously used for a different measurement, is located below the Pt1000 sensor in the position normally used for measurements with current flow.
	\protect\subref{fig:EmissivityCalibration_Photos} Photos of the Pt1000 sensor with the \ce{Ca2RuO4} sample used to calibrate the emissivity.
	The circles indicate two areas used to extract $\Tr$ for the sample (in orange) and for the reference material (alumina, substrate of the Pt1000 sensor, in green).
	The orange circle is carefully chosen in order to select an area of the sample where minimal damage occurs during the several thermal cycles, as can be seen comparing the images for cycle 1 and 7.
	\protect\subref{fig:EmissivityCalibration_SampleT} Raw infrared reading for the sample during several temperature cycles performed to assess the thermal contribution of the surrounding areas.
	In the first 3 cycles $\Tp$ is fixed to 0, 25, and \SI{60}{\celsius} respectively, while the sample temperature is varied by passing a current through the Pt1000 sensor.
	This configuration keeps most of the camera field of view at a fixed temperature value, while only a small region around the sample area is heated up.
	In the last measurement, the temperature of both the sample and the Pt1000 is controlled directly by the Peltier stage ($\Tp=\Tsm$), while only a small current is sourced to the Pt1000 to read its temperature.
	In this configuration, the whole background is also heated up, so that the whole area in the camera field of view experiences a change in temperature.
	Due to this increased background signal, we measure higher values of $\Tr$.
	Since in all 4 configurations the sample is in thermal equilibrium with the Pt1000, the dependence of $\Tr$ on $\Tp$ indicates that the background has an important effect on the thermal reading of the camera.
	It is important to take this effect into account to properly estimate the sample temperature.
	To obviate this issue, the emissivity calibration is performed as described in the following section.
	\protect\subref{fig:EmissivityCalibration_AluminaT}Raw reading for the alumina during the same cycles.
	A similar increasing trend of $\Tr$ with $\Tp$ is observed.
	However, the temperature increase is smaller since the material has a higher emissivity (larger slope) that determines an increased instrumental sensitivity.
	\protect\subref{fig:EmissivityCalibration_SampleCycle} Sample and
	\protect\subref{fig:EmissivityCalibration_AluminaCycle} alumina trend of $\Tr$ during repeated cycles performed with the same conditions.
	For this reproducibility test, we set $\Tp=\SI{25}{\celsius}$ and control the sample temperature by sourcing a current through the Pt1000 sensor.
	The curves for the alumina in \protect\subref{fig:EmissivityCalibration_AluminaCycle} are overlapping, indicating a good reproducibility and stability of the camera reading.
	The curves for the sample in \protect\subref{fig:EmissivityCalibration_SampleCycle}, instead, show some variability which we link to a change in sample conditions.
	This can be related to the formation of macroscopic cracks that affect the sample surface and may change its tilt angle.
	The variability is larger at higher temperatures, especially for the metallic phase which has a smaller slope.
	For these reproducibility considerations, for the uncertainty in the linear regressions discussed in \cref{fig:TIR_Calibration}, and also for the unavoidable sample-to-sample variations, we estimate an error of about \SI{\pm10}{\celsius} on the calibrated values of $\Tins$ and $\Tmet$.
	}
	
	\label{fig:EmissivityCalibration}
	
\end{figure*}

\begin{figure*}[h]
	\includegraphics[page=9,width=140mm]{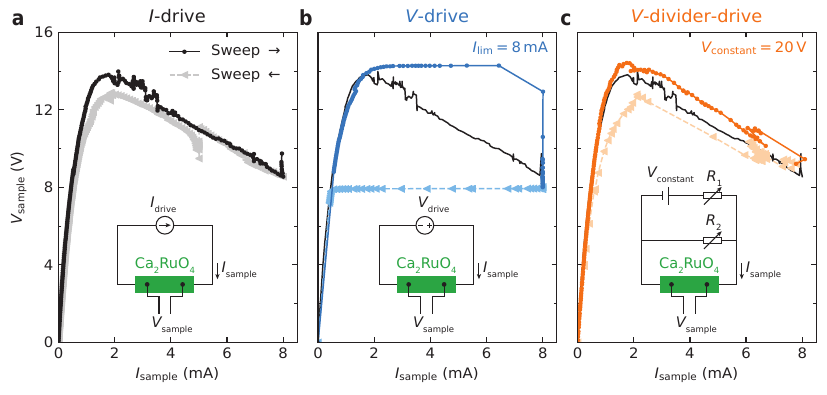}
	
	\subfloat{\label{fig:DriveCurrent}}
	\subfloat{\label{fig:DriveVoltage}}
	\subfloat{\label{fig:DriveVdivider}}

	\caption{\textbf{Different configurations to measure the voltage--current characteristics.}
		\protect\subref{fig:DriveCurrent} Current drive, where the sample is directly connected to the output of a variable current supply (Sample \#1, CR29c1a).
		This configuration allows to measure all the points on the experimental curve and it is the one used in the main text.
		The increasing-current sweep is represented by the solid colour (black, repeated in the following graphs for comparison purposes), while the decreasing-current sweep is in the shaded colour (grey, same convention for the following graphs).
		\protect\subref{fig:DriveVoltage} Voltage drive, where a variable voltage source with a current limit of \SI{8}{\milli\ampere} is used.
		This configuration allows to measure only a part of the experimental points and is unstable above the voltage peak of the $V$--$I$ curve, where the current tends to increase very quickly while the current source reacts to comply with the current limit.
		For this reason, upon exceeding \SI{14}{\volt} the current jumps to the compliance value of \SI{8}{\milli\ampere}; similarly, upon decreasing the voltage, the current jumps back to the original characteristics (in black) at the point corresponding to \SI{8}{\volt} and \SI{0.5}{\milli\ampere}.
		This configuration is similar to what has been used in [Nakamura F. \textit{et al.}, Sci. Rep. \textbf{3}, 2536 (2013)].
		\protect\subref{fig:DriveVdivider} Voltage-divider drive, where a constant voltage source of \SI{20}{\volt} is connected in a circuit with two variable resistors (up to \SI{10}{\kilo\ohm} each) that are used to manually control the current through the sample.
		This configuration avoids large current or voltage spikes through the sample and allows to acquire the complete $V$-$I$ characteristics.
		This configuration is similar to what has been used by [Zhang J. \textit{et al.}, Phys. Rev. X \textbf{9}, 011032 (2019)].
		In all configurations, the current flowing through the sample $I_\mathrm{sample}$ and the four-probe voltage $V_\mathrm{sample}$ are measured independently.
		Apart from the missing or unstable points, the three experimental configurations uncover the same $V$--$I$ characteristics of \ce{Ca2RuO4} at room temperature.
	}
	
	\label{fig:DifferentDrives}
	
\end{figure*}

\begin{figure*}[h]
	\includegraphics[page=10,width=120mm]{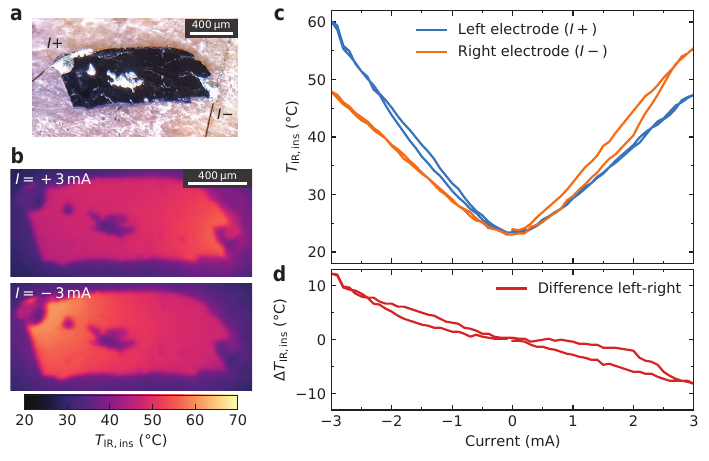}
	
	\subfloat{\label{fig:Peltier_photo}}
	\subfloat{\label{fig:Peltier_IR}}
	\subfloat{\label{fig:Peltier_T}}
	\subfloat{\label{fig:Peltier_dT}}
	
	\caption{\textbf{Current-induced Peltier heating of \ce{Ca2RuO4}.}
		\protect\subref{fig:Peltier_photo} Photograph of the single crystal used for the experiment (Sample \#9, CR29G1).
		Electrical contact is provided by two gold wires (\SI{18}{\micro\metre} in diameter) directly connected to the sample surface and edges by Ag epoxy, without any gold plating.
		To minimise the thermal conductivity with the bath at $\Tp=\SI{25}{\celsius}$, the sample is resting on cigarette paper without the use of any glue, and the cigarette paper is lifted by about \SI{2}{\milli\metre} from the underlying sample holder.
		\protect\subref{fig:Peltier_IR} Thermal images plotted with the insulating calibration $\Tins$ upon driving a positive and negative current through the sample.
		The material is kept in the insulating phase for the whole experiment.
		\protect\subref{fig:Peltier_T} Thermal temperature close to the left and right electrodes as a function of current.
		\protect\subref{fig:Peltier_dT} Difference in temperature between the two electrodes.
	}
	
	\label{fig:PeltierHeating}
	
\end{figure*}

\else
\fi

\end{document}